\documentclass[twocolumn,runningheads,natbib]{svjour2}
\smartqed  
\usepackage{graphicx}
\journalname{Astrophysics and Space Science}
\begin{document}

\title{High frequency oscillations during magnetar flares
}
\subtitle{Evidence for neutron star vibrations}


\author{Anna L. Watts \and Tod E. Strohmayer}

\authorrunning{Watts \& Strohmayer} 

\institute{A.L. Watts \at
              Max Planck Institut f\"ur Astrophysik, Karl-Schwarzschild-Str. 1, 85741 Garching, Germany \\
              Tel.: +49-89-300002015\\
              Fax: +49-89-300002235\\
              \email{anna@mpa-garching.mpg.de}           
           \and
           T.E. Strohmayer \at
              NASA Goddard Space Flight Center, Exploration of the Universe Division, Mail Code 662, Greenbelt, MD 20771, USA
}

\date{Received: date / Accepted: date}

\maketitle

\begin{abstract}
The recent discovery of high frequency oscillations during giant flares from the Soft Gamma Repeaters SGR 1806-20 and SGR 1900+14 may be the first direct detection of vibrations in a neutron star crust.  If this interpretation is correct it offers a novel means of testing the neutron star equation of state, crustal breaking strain, and magnetic field configuration. We review the observational data on the magnetar oscillations, including new timing analysis of the SGR 1806-20 giant flare using data from the {\it Ramaty High Energy Solar Spectroscopic Imager} and the {\it Rossi X-ray Timing Explorer}.  We discuss the implications for the study of neutron star structure and crust thickness, and outline areas for future investigation. 

\end{abstract}

\section{Introduction}
\label{intro}
The Soft Gamma Repeaters (SGRs) are thought to be magnetars, neutron stars with magnetic fields in excess of $10^{14}$G \citep{duncan1992, thompson1995}. Decay of the strong field powers regular gamma-ray flaring activity that culminates, on rare occasions, in a giant flare with a peak luminosity in the range $10^{44}$- $10^{46}$ erg~s$^{-1}$.  The three giant flares detected to date consist of a short, spectrally hard initial peak, followed by a softer decaying tail that lasts for several hundred seconds.  Pulsations with periods of a few seconds are visible in the tail and reveal the neutron star spin.  Their presence is thought to be due to a fireball of ejected plasma, trapped near the stellar surface by the strong magnetic field \citep{thompson1995}.   

Powering the giant flares requires a catastrophic global reconfiguration of the magnetic field.  The coupling between the field and the charged particles in the neutron star crust means that this is likely to be associated with large-scale crust fracturing \citep{flowers1977, thompson1995, thompson2001, schwartz2005}.  This in turn should excite global seismic vibrations:  on Earth seismologists regularly observe global modes after large earthquakes (see for example \citealt{park2005}).  In the SGR case, the coupling of field and crust should cause the modes to modulate the X-ray lightcurve.  

Various different types of oscillation are possible, but theory suggests that the easiest to excite and observe should be the toroidal shear modes of the crust \citep{blaes1989}.  The particular harmonics excited will depend on the size, shape and speed of the fracture.  The observed mode frequencies depend on the neutron star mass and radius (via gravitational redshift and the influence of the equation of state on crust structure), crustal composition, and magnetic field strength and configuration \citep{hansen1980, schumaker1983, mcdermott1988, strohmayer1991, duncan1998, messios2001, piro2005}.  Detection and identification of crustal modes would therefore probe all of these areas of neutron star physics.

In 2004, SGR 1806-20 emitted the most powerful flare ever recorded.  In a landmark paper, \cite{israel2005} reported the detection of Quasi-Periodic Oscillations (QPOs) in the 18-93 Hz range in the decaying tail of the flare.  Prompted by this result, \cite{strohmayer2005} re-analysed the 1998 giant flare from SGR 1900+14 and found QPOs in the range 28-155 Hz. In both cases tentative identifications can be made with a sequence of toroidal modes.  More in-depth analysis of the SGR 1806-20 flare has since revealed additional QPOs, including the first evidence for a higher radial overtone \citep{watts2006, strohmayer2006}.  We review all of the observational results in Section \ref{sec:1}.   In Section \ref{sec:2} we discuss the results in the context of the toroidal mode models, and show how they can be used to constrain neutron star parameters including the crust thickness.  We conclude in Sections \ref{sec:3} and \ref{sec:4} with a discussion of outstanding issues.  

\section{Observational results}
\label{sec:1}

\subsection{SGR 1806-20}
\label{sec:1a}

The December 2004 flare was the brightest ever recorded, with a peak luminosity of $\sim 10^{46}$ erg s$^{-1}$.  Analysis of data from the {\it Rossi X-ray Timing Explorer} (RXTE) by \cite{israel2005} revealed a transient 92.5 Hz QPO in the tail of the flare, associated with a particular rotational phase.  The QPO appeared around 170s after the main peak, at the same time as an apparent boost in unpulsed emission. The presence of weaker 18 and 30 Hz features at late times was also suggested.

The toroidal modes are labelled by the standard quantum numbers $l$, $m$ and $n$, the first two being angular quantum numbers, the final one denoting the number of radial nodes.  The 30 and 92.5 Hz features found by \cite{israel2005} can be identified as the fundamental $l=2, n=0$ mode and the $l=7, n=0$ harmonic (see Section \ref{sec:2} for a discussion of current mode models).  The 18 Hz feature, by contrast, is at too low a frequency to be identified as a toroidal mode of the crust.  

The other spacecraft with high time resolution data of the SGR 1806-20 giant flare was the {\it Ramaty High Energy Solar Spectroscopic Imager} (RHESSI), a solar-pointing satellite that covers a wider energy band than RXTE.  RHESSI's detectors are split into front and rear segments.  Strong albedo flux from the Earth affected the time resolution of the rear segments (making them unsuitable for studying frequencies $>$ 50 Hz), but when the rear segments are included the countrate recorded by RHESSI exceeded that of RXTE.  

\cite{watts2006} analysed the RHESSI data of the flare and confirmed the presence of the 92.5 Hz QPO, at the same time and rotational phase found by \cite{israel2005}.   At low frequencies, where the larger RHESSI countrate gave added sensitivity, broad QPOs at 18 and 26 Hz were found 50-200 s after the main flare, at the same rotational phase as the 92.5 Hz QPO.  Although there was a weaker feature at 30 Hz, it was not at the 3$\sigma$ level after accounting for the number of trials so we were not able to make a robust confirmation of the \cite{israel2005} result.  Subsequent closer comparison of the two datasets, however, indicates that RHESSI's countrate is not the only factor affecting its sensitivity.  As such we now believe that the 30 Hz detection is robust (see below).  

Also detected in the RHESSI data was a 625 Hz QPO\footnote{Grid shadowing in RHESSI can give rise to spurious high frequency signals, and there is a known artifact at 718 Hz in the giant flare data (Gordon Hurford, private communication).  The phenomenon affects certain detectors far more strongly than others since it is due to spacecraft geometry.  It is also (perhaps unsurprisingly) more prominent at lower energies. The fact that the 625 Hz QPO is dependent on the rotational phase of the magnetar, despite there being comparable numbers of photons at other rotational phases, is a good indication that it is not a detector artifact. In addition it is only detected at higher energies, where shadowing should be less of an issue.  However, to be sure we also tested the variation of QPO power when photons from individual detectors were removed from the analysis. The drops in power were consistent with the drop in countrate, confirming that the signal is not anomalously strong in one or two detectors.  As such we are confident that the signal is indeed associated with the source.} at an energy band nominally higher than that recorded by RXTE\footnote{Direct comparison of recorded energies is difficult because for both spacecraft the flare was off-axis, resulting in substantial scattering within the body of the satellite.}.  A full discussion of the QPO properties and the significance of the detection can be found in Section 2 of \cite{watts2006}.  Compared with the 92.5 Hz QPO the 625 Hz QPO is in a higher energy band, emerges earlier in the tail of the flare, and is at a different rotational phase.  Excitingly, however, it is also at the approximate frequency expected for the radial overtone $n=1$ toroidal modes \citep{piro2005}. 

We have recently completed a more in-depth analysis of the now public RXTE dataset \citep{strohmayer2006}.  We start by discussing the low frequency QPOs.  In this regime the RXTE countrates are lower than RHESSI's.  However, at times when the 18 and 26 Hz QPOs are particularly strong in the RHESSI dataset they are also detected in the RXTE data.  We are thus able to confirm both the frequencies and the rotational phase dependence.  Further analysis of the 30 Hz QPO confirms the detection made by \cite{israel2005}, and reveals that this QPO is also strongly rotational phase dependent.  Rather surprisingly, given the difference in countrates, this feature is far stronger in the RXTE data than in the RHESSI dataset.  There are several factors, however, that could account for this discrepancy (see the discussion in \citealt{strohmayer2006}).  

Looking to higher frequencies, a rotational phase dependent search reveals additional QPOs at 150 Hz, 1840 Hz and, most excitingly, at 625 Hz (the latter appearing at the same rotational phase as the 92.5 Hz QPO).  Full details of the QPO properties and the significance of the detection can be found in Section 2.1 of \cite{strohmayer2006}.  There are, it turns out, intriguing differences between the RHESSI and RXTE 625 Hz QPOs.  The RXTE QPO has lower fractional amplitude, lower coherence,  a different rotational phase association, and appears at later times in a lower energy band.  One possibility is that we are seeing time evolution of one QPO whose photon energy and amplitude are falling over the course of the tail.  The second possibility is that we are seeing two different modes.  As pointed out by \cite{piro2005} the $n=1$ modes are nearly degenerate in $l$, so there are several different modes with very similar frequencies.  Despite the differences, it seems extremely unlikely that two independent instruments would detect signals at a consistent frequency unless that frequency were intrinsic to the source.

\subsection{SGR 1900+14}
\label{sec:1b}

In August 1998 SGR 1900+14 emitted a giant flare with a peak luminosity $\sim 10^{44}$ erg s$^{-1}$.  The flare was detected by RXTE, albeit with data gaps due to the configuration of the spacecraft at that time.  \cite{strohmayer2005} started by analysing each good interval separately, and discovered a strong transient 84 Hz QPO during a $\sim 1$ s interval about 60 s after the main flare.  Folding up data from more intervals revealed a pair of persistent QPOs at 53.5 and 155.5 Hz, with a weaker feature at 28 Hz, all at the same rotational phase as the 84 Hz signal.  No QPOs were detected at other rotational phases. The scaling of the QPO frequencies is what the existing models suggest for the $n$=0, $l$=2, 4, 7 and 13 toroidal modes of the crust.   

\section{Constraining neutron star properties}
\label{sec:2}

Detection of signals at similar frequencies in the giant flares from two different SGRs implies that the same process is operating in both objects.  In addition, the strong rotational phase dependence of {\it all} of the detected QPOs provides strong evidence that the modulations are associated with the stellar surface.  In this section we discuss the implications of the QPOs in the light of the toroidal mode model.  Existing models do have shortcomings, which we discuss in more detail in Section \ref{sec:3}.  However, crust mode models remain at present the most promising mechanism and we will proceed accordingly.

For a non-rotating, nonmagnetic star, \cite{duncan1998} estimates the frequency $\nu$ of the fundamental $l=2, n=0$ toroidal mode (denoted $_2t_0$) to be

\begin{equation}
\label{eqn1}
\nu(_2t_0) = 29.8 ~ \frac{(1.71 - 0.71 M_{1.4} R_{10}^{-2})^{1/2}}{0.87 R_{10} + 0.13 M_{1.4} R_{10}^{-1}} \mathrm{~~~Hz}
\end{equation}
where $R_{10} = R/10$ km and $M_{1.4} = M/1.4 M_\odot$.  The frequencies of the higher order $n=0$ modes are given by

\begin{equation}
\label{eqn2}
\nu(_lt_0) = \nu(_2t_0) \left[ \frac{l(l+1)}{6}\right]^{1/2} \left[ 1 + \left(\frac{B}{B_\mu}\right)^2\right]^{1/2}
\end{equation}
where the final factor is a magnetic correction, $B_\mu \approx 4\times 10^{15} \rho_{14}^{0.4}$ G and $\rho_{14} \sim 1$ is the density in the deep crust in units of $10^{14}$ g cm$^{-3}$.  In deriving this equation it is assumed that magnetic tension boosts the field isotropically.  Field configuration and corresponding non-isotropic effects could however alter this correction dramatically \citep{messios2001}.  More recent calculations by \cite{piro2005} using better models of the neutron star crust confirm these frequency estimates.  

\begin{figure}
\centering
\includegraphics[clip, width=8cm]{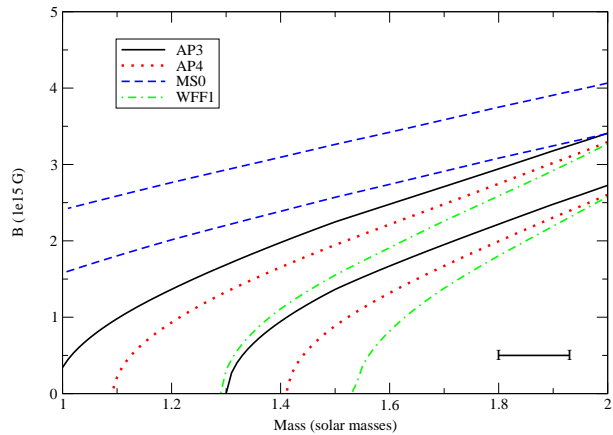}
\caption{Mass and magnetic field required to give the toroidal mode frequencies inferred for SGR 1806-20 and SGR 1900+14.  We show results for four of the EOS from \cite{lattimer2001}; in order of increasing stiffness they are WFF1, AP4, AP3 and MS0.  The upper line for each EOS indicates the parameters necessary to give the SGR 1806-20 frequencies; the lower line those for the lower SGR 1900+14 frequency.  The horizontal line indicates the uncertainty in the position of the footpoints of the lines due to the width of the QPOs.}
\label{fig1}      
\end{figure}

The detection of a set of modes from both SGR 1806-20 and SGR 1900+14 with the expected  $[l(l+1)]^{1/2}$ scaling in frequency is indicative of the presence of toroidal modes.  We can then ask what can be learnt from the fact that the fundamental $_2t_0$ mode frequency seems to differ for the two SGRs.  For SGR 1806-20 it is inferred to be 30 Hz, whereas for SGR 1900+14 it is lower, at 28 Hz. For this to be the case, the properties of the two stars must differ.  Given an equation of state (EOS), equations (\ref{eqn1}-\ref{eqn2}) specify the relationship between mass and magnetic field necessary to give modes at the inferred frequencies.  Figure \ref{fig1}, taken from \cite{strohmayer2005}, shows for several EOS the stellar parameters that give $_2t_0$ oscillations at the frequencies inferred for the two stars.  

Several conclusions follow from Figure \ref{fig1}.  Firstly, unless the masses differ substantially, SGR 1806-20 must have a stronger magnetic field than SGR 1900+14.  This accords with estimates of field strength from timing studies \citep{woods2002}.  Secondly, the highest masses and the hardest EOS require fields far higher than those inferred from timing, whereas the softest EOS require high masses.  Clearly these inferences are all model-dependent, and could change as the models get more sophisicated.  However, it illustrates the potential of neutron star seismology to constrain the equation of state.  

If the 625 Hz QPOs detected in the SGR 1806-20 flare are indeed $n=1$ toroidal crust modes then the implications are profound, since from the ratio of frequencies of the $n=0$ and $n=1$ modes one can deduce the thickness of the crust, $\Delta R$ (gravitational redshift factors scale out). In the limit of a thin crust ($\Delta R \ll R$) and constant shear wave speed \citep{hansen1980, mcdermott1988, piro2005} one can show that

\begin{equation}
\frac{\Delta R}{R} = \left(\frac{3}{2}\right)^{1/2} \frac{\nu(_2t_0)}{\nu(_lt_1)}
\end{equation}
So given $\nu(_2t_0) = 30$ Hz and $\nu(_lt_1) = 625$ Hz we find $\Delta R/R = 0.06$.  A more sophisticated estimate that does not assume a thin crust or a constant shear speed gives $\Delta R/R$ in the range 0.1 - 0.12 \citep{strohmayer2006}.

Measuring the crust thickness also gives an additional constraint on the EOS (independent of redshift effects), since stellar models of a given mass computed with different EOS will in general have crusts with different depths.  The article by Lattimer in this volume outlines how this could be used to constrain the nuclear symmetry energy and the nuclear force model.   

\section{Theoretical issues}
\label{sec:3}

Neutron star seismology, revealed during giant flares from magnetars, has great potential as a probe of stellar structure, composition and magnetic field geometry.  There are many theoretical issues, however, that remain to be resolved.  One of the main issues is that of the coupling between the crust and the fluid core due to the presence of the strong magnetic field.  Global magneto-elastic modes of the whole star may be necessary to explain, in particular, the QPOs at 18 and 26 Hz, which do not sit comfortably within existing toroidal mode models.  The coupling between the crust and the field will also determine how the modes modulate the lightcurve. 

Modes of the neutron star crust (most likely coupled to the core via the magnetic field) remain the most promising mechanism identified to date for the QPOs.  Although several other suggestions have been made, all have serious difficulties.Modes of the magnetosphere, mentioned by \cite{levin2006}, are likely to have too high a frequency due to the high Alfv\'en speed in this region.  Modes of the trapped fireball are unlikely since one would expect the frequency to change as the fireball shrinks:  no such correlation is observed.   The third possibility, raised by Alpar at this meeting, is interaction with a debris disk and a mechanism similar to that thought to generate QPOs in the accreting systems.  However even if a small disk were present in these systems, it would be difficult to explain the rotational phase dependence.   

The complex temporal variation of the QPOs also requires explanation.  In both giant flares some of the QPOs seem to be highly transient, whereas others persist for most of the flare.  We mentioned in Section \ref{sec:2} the possible evolution of the 625 Hz QPO seen during the SGR 1806-20 flare.  However the 92.5 Hz QPO also shows variation in frequency and amplitude \citep{israel2005, strohmayer2006}.  Evolution of the magnetic field in the aftermath of the flare, relaxation of the deep crust, or evolution of the surrounding plasma are all candidates for causing such variations.  Damping and excitation mechanisms are a critical area for future study.  

\section{Conclusions}
\label{sec:4}

The discovery of high frequency oscillations during giant flares from the Soft Gamma Repeaters SGR 1806-20 and SGR 1900+14 may be the first direct detection of vibrations in a neutron star crust.  If this interpretation is correct it offers a new means of testing the neutron star equation of state, crustal breaking strain, and magnetic field configuration.  In particular, if the mode interpretation is secure, it allows us to make the first direct estimate of the thickness of a neutron star crust.  This is particularly impressive when one considers the fact that all of the data acquired so far have been obtained by chance using satellites that were observing other objects.  If a rapid-slew instrument such as SWIFT could be configured to record high time resolution data throughout the tails of giant flares from known SGRs, the potential for additional discoveries is immense.

\begin{acknowledgements}
ALW acknowledges support from the European Union FP5 Research Training Network `Gamma-Ray Bursts:  An Enigma and a Tool'.  TES thanks NASA for its support of high energy astrophysics research.
\end{acknowledgements}

\end{document}